\documentclass[twocolumn]{aastex62}
\usepackage{graphicx}
\usepackage{epstopdf}
\usepackage{afterpage}

\shorttitle{AstroSat UVIT M31 Source Catalog}
\shortauthors{Leahy et al.}

\begin{document}


\title{AstroSat UVIT Survey of M31: New Compact Source Catalog}


\author{Denis Leahy, Sujith Ranasinghe, Saptadwipa Mitra, Kripansh Rawal}
\affil{Dept. of Physics $\&$ Astronomy, University of Calgary, University of Calgary,
Calgary, Alberta, Canada T2N 1N4}



\begin{abstract}
An ultraviolet survey of M31 has been carried out during 2017-23 with the UVIT instrument onboard the AstroSat Observatory.
Here we present far and near ultraviolet (FUV and NUV) observations from the M31 UVIT survey,
which covers a sky area of $\simeq  3.5^\circ \times 1.3^\circ$  with spatial resolution of $\simeq1^{\prime \prime}$. 
The observations included six filter bands in the wavelength range of 120 nm to 280 nm.
Including the six bands, $\simeq$115,000 sources with signal-to-noise S/N$\ge$3 ( $\simeq$95,000 sources with signal-to-noise S/N$\ge$5) were detected at FUV or NUV wavelengths, with the largest set of detections ($\simeq$54,000 sources) in the FUV 150 nm band  (F148W filter). 
This is considerably more than for the first version of the M31 source catalog (published in 2020), in part due to additional observations of M31 by UVIT and in part due to improved data processing.
The magnitude (m$_{AB}$) at which incompleteness sets in is $\simeq$23.0 in the F148W band, with the other bands somewhat less sensitive, with least sensitive band (in m$_{AB}$ units) N279N with incompleteness for sources fainter than $\simeq$20.3).
The faintest sources detectable in F148W have m$_{AB}\simeq$25.4, with other bands having higher minimum detectable brightness, with N279N having minimum detectable limit of m$_{AB}\simeq$22.1.
The product of this work is the new M31 UVIT compact source catalog, containing positions, fluxes , magnitudes and S/N for the sources. 

\end{abstract}


\keywords{}



\section{Introduction}\label{sec:intro}

M31 is the most prominent nearby large galaxy to the Milky Way and is at distance 783 kpc \citep{2005McConnachie}.
Thus the objects in M31 have nearly uniform distances, giving M31 studies an advantage over studies of stars, supernova remnants and X-ray sources in our own Galaxy, which often have uncertain distances.

M31 has been studied thoroughly at optical wavelengths, e.g., \citet{1986Massey}, \citet{1993Davidge}, and \citet{2001Ibata}. 
The Hubble Space Telescope (HST) has carried out a survey of part of M31 \citep{2012Bianchi} in 6 wavebands 
covering $\sim9$ kpc$^2$ of a total area of M31 of $\sim300$ kpc$^2$.
A significantly larger area in the central and northeastern parts of M31 ($\sim$1/4 of the galaxy) was covered by the Panchromatic Hubble Andromeda Treasury survey (PHAT,  \citealt{2014Williams}), which covered 6 bands from 275 nm to 1600 nm and cataloged 117 million stars.
These two HST surveys included ultraviolet, where hot-massive stars strongly emit \citep{2014Bianchi}
and are well suited to studying hot stars in M31.

GALEX \citep{2007Morrissey} observed M31  \citep{2005Thilker}  in FUV (154 nm) and NUV (232 nm) bands and had $\sim5^{\prime \prime}$ spatial resolution. 
With this data 894 star-formation (SF) regions were detected in M31 \citep{2009Kang}. 
The GALEX and optical data were further analyzed to produce a catalog of 700 star clusters in M31 \citep{2012Kang}. 
M31 was imaged in NUV by the Swift/UVOT instrument using 3 filters, with central wavelengths of 194, 225 and 260 nm, filter widths of 51 to 65 nm, spatial resolutions of 2.4 to 2.9${^\prime\prime}$ and exposures of 1450 to 1680 s (Chapter 5 of \citealt{2017PhDT.......221H}). 
The 3 UVOT bands are combined with SDSS optical and Spitzer infrared data, to carry out a study of dust properties and star formation for M31.  
The UVOT results for M31 are summarized in the review paper of \cite{2024Univ...10..330S}.

UVIT (UltraViolet Imaging Telescope) is one of the main instruments on the AstroSat mission \citep{singh+2014}, together with the three X-ray instruments: SXT, LAXPC and CZTI. 
The UVIT telescope, with its high spatial resolution of $\simeq1^{\prime \prime}$, and its calibration are described in \citet{2017Tandon2},  \citet{2011Postma} and references therein.
UVIT can resolve stellar clusters and a large number of individual stars in M31, and the NUV and FUV filters are ideal for studying hot stars in M31. 

A catalog of M31 UVIT sources was published by \cite{2020ApJS..247...47L} using the 2 years of the survey, from Oct. 10, 2016 to Nov. 25, 2018. 
That catalog included 18 different fields in several FUV and NUV filter bands and resulted in $\sim$31,000 F148W sources, $\sim$1900 F154W sources, $\sim$11,000 F169M sources, $\sim$6700 F72M sources, $\sim$15,000 N219M sources and $\sim$8600 N219M sources. 
\cite{2021AJ....162..199L} discovered FUV variable sources in M31 and showed that they were mostly associated with star clusters, HII regions, semi-regular variables and novae.
Studies of star clusters in the northern disk of M31 with UVIT were carried out by \cite{2022AJ....164..183L}, showing that there were younger star clusters than previously found.
The complex structure of the M31 bulge, including its boxiness and multiple components was revealed by analysing the UVIT FUV image of the M31 bulge \citep{2023ApJS..265....6L}.
UVIT was  also used to discover 20 UV-emitting supernova remnants in M31 \cite{2023AJ....165..116L}. 
\cite{2024AJ....167..211L} carried out multiband photometry in 78 regions covering M31 to measure the star formation history of M31, was sensitive to and measured the spatial variation) of the youngest stellar populations.

Since 2018 a similar amount of observing has been added to the survey, and the instrument calibration and data processing methods have significantly improved, including updated instrument calibration \citep{2020AJ....159..158T}, more accurate coordinate matching to position calibrators \citep{2020PASP..132e4503P}, improved data reduction software \citep{2021JApA...42...30P}, optimization of pointing by correcting for spacecraft bus jitter \citep{2023JApA...44...12P}, and methods for improved source extraction in crowded fields (appendix in 
\citealt{2021AJ....161..215L}). Finally, the availability of the Gaia DR3 source catalog as an astrometric reference \citep{2023A&A...674A...1G} has allowed more accurate position calibration of the UVIT data.
The UVIT data reduction and analysis software packages are now in their final form because the mission is well past its 5-year planned lifetime. 
Thus, now it is timely to produce the new UVIT M31 Survey compact source catalog. 

This paper presents the new compact source catalog derived from the full AstroSat UVIT M31 survey project, with updated calibration, astrometry and data processing. 
The main goal of the M31 survey was to obtain FUV and NUV imaging and photometry for M31 utilizing the high spatial resolution of the UVIT instrument on the Astrosat satellite. 
M31 was observed, over a period of 2016 through 2024, with a set of 23 pointings, each with 1 to 5 observations in different filters, to cover the galaxy. 
The sky area coverage is about 3.5$^\circ$ (along the major axis of M31) by 1.3$^\circ$ (along the minor axis) with a spatial resolution of $\simeq1^{\prime \prime}$.
The primary product of the current work is the new M31 UVIT compact source catalog. 
In Section \ref{sec:obs}, we describe the observations and data reduction, including astrometry and photometry. 
We present the M31 UVIT source catalog and summarize the catalog in Section \ref{sec:results}.

\section{Observations and Analysis}  \label{sec:obs}

The UVIT instrument has two 38 cm telescopes both with circular field-of-view with diameter  $\simeq$28$^{\prime}$.
One telescope is for FUV and one is for NUV and visible (VIS) wavelengths, with a dichroic that splits the light into NUV and VIS.  
There are several filters with different bandpasses, as described in \citet{2017Tandon2}.
The VIS channel provides pointing information and is not used for photometry.
The full description of UVIT is given in \citet{2016Subra}, \citet{2017Tandon2}, \citet{2017Tandon1} and \citet{2020AJ....159..158T}.

UVIT's typical astrometric errors of less than 0.5 $^{\prime \prime}$\citep{2017Tandon2} have been further improved by using astrometric calibration software \citep{2020PASP..132e4503P} which follows the CD-Matrix standard for the Gnomic projection as discussed in \citet{2002Calabretta}. 
This is implemented in CCDLAB \citep{2021JApA...42...30P}  and matches approximately one-hundred sources per image, currently utilizing the Gaia DR3 catalog for position calibration.
We also calculate better pointing corrections for UVIT \citep{2023JApA...44...12P}. 
The current source position errors are measured by the root-mean-square (RMS) in offset of the UVIT sources cross-matched with Gaia DR3 sources. The RMS offsets have typical values of 0.2 to 0.25 $^{\prime \prime}$for our fields. 

\begin{figure}
\centering
\epsscale{1.0}
\plotone{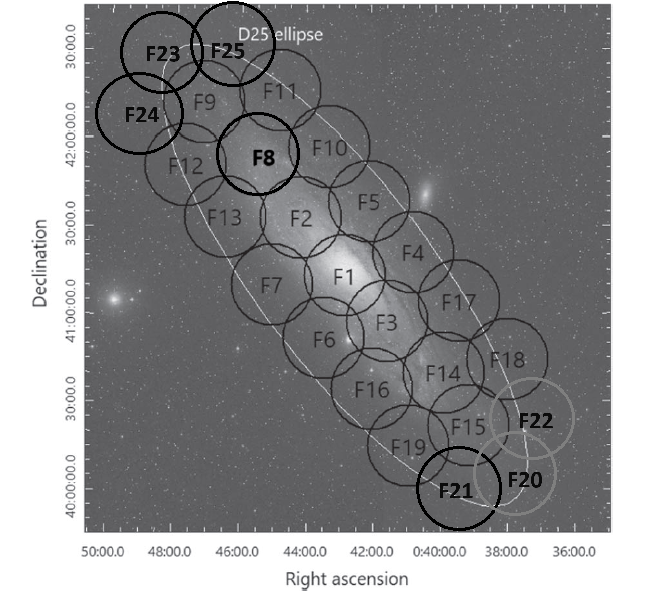}
\caption{The  DSS Poss2 blue filter image of M31 (greyscale image). Right ascension and declination are J2000 coordinates. The white ellipse shows the D25 ellipse for M31 \citep{2007ApJS..173..185G}. The
locations of the fields for the UVIT survey of M31 are shown by the circles. F8, F21 and F23 to F25 are new fields marked with thicker black circles. Fields F20 and F22 (grey circles) were not observed because of stars in those fields above the brightness limit for instrument safety.
 \label{fig:fieldpositions}
}
\end{figure}

The M31 UVIT survey used the FUV filters F148W, F154W, F169M and F172M and the NUV filters N219M and N279N. 
The whole area of M31 (except Field 21) was covered by the most sensitive (and broadest band) FUV F148W filter, whereas the other filters only covered parts of M31.
The sky locations in RA, Dec (J2000, decimal degrees) of the field positions for the UVIT survey of  M31
are shown in Figure~\ref{fig:fieldpositions} overlaid on the DSS Poss2 blue filter image of M31.
Table~\ref{table:obs} lists the positions of the centers for observed Fields 1 to 19, 21, and 23 to 25, and lists the different filters used for each field. 
Fields 20 and 22 were not observed because of bright stars above the safe limit for the instrument.
Exposure times are given in columns 5 and observation dates (expressed as solar-system Barycentred Julian Date) for all field/filter combinations are given in column 6. 
Some filter bands were observed multiple times: these are listed with subscripts and exposure times and dates are given for each observation.
The multiple images of the same field and same filter were combined into a single image prior to source finding, with net exposure times for the combined images given in footnote d of Table 1.

\subsection{Source Extraction and Source Photometry}\label{sec:extraction}

Source finding was carried out using CCDLAB for each image with different field and filter, i.e. for a total of 68 different images.  
Photon counting noise is large for sources with low number of detected counts, such as around the field edges where the exposure time is low or for faint sources.
To improve the source detection reliability we convolved each image with a Gaussian smoothing function, $\sigma$=1.5 pixels (0.63 arcsec), prior to source extraction \citep{2020ApJS..247...47L}. 

We set the source extraction parameters in CCDLAB to optimize detection for the main part of the field (all but the field edges).
This resulted in a ring of false sources (seen as higher source density) around the edge (outer $\sim1^{\prime}$) of each circular field.
The ring is caused by higher noise caused by the low exposure time at the image edge.
We eliminated this outer $\sim1 ^{\prime}$ ring as follows:
the size of the ring for each image was determined by carrying out an initial source extraction;  
the initial source map was used to choose the circular area of the image (call the Region Of Interest or ROI) which avoided the outer ring of  sources; 
the source extraction was repeated using only the area of the ROI to produce a source list. 
Table~\ref{table:roi} gives the center and radius for each ROI for the different field/filter combinations. 

The next step was determining source counts for each detected source. 
The UVIT instrument was calibrated using isolated point sources \citep{2017Tandon2}, for which a curve-of-growth (COG) analysis was used to extract net counts. 
A significant fraction of the M31 sources are not well-isolated, and are instead often separated by less than 
 the size of the wings of the point-spread function (PSF), which extend to $\sim11^{\prime \prime}$ (Fig. 17 of  \citealt{2017Tandon2}).
For the M31 sources, which are not well isolated, we found an elliptical Gaussian 
gave more consistent photometry than other fit functions (section 2.3 of  \citealt{2020ApJS..247...47L}).
Thus, we fit elliptical Gaussian functions with a fixed box size to the sources in the UVIT images. 
Because the Gaussian fits the core of the PSF, a multiplicative correction is needed to obtain total counts from any given source including those in the PSF wings.
The correction factor was found by fitting isolated point sources to determine the ratio of counts from curve-of-growth (COG) fits to counts from elliptical Gaussian fits. 
This ratio had a mean value of 1.82 \citep{2020ApJS..247...47L}, which was verified for the current images. 

The next step was to obtain source flux and its uncertainty from source counts. 
Count rate was obtained by dividing corrected counts by the exposure time for each image. 
The source count rates were corrected for coincidence loss.
 The brightest $\sim$70 sources (in F148W filter) and 5-20 sources (in the other filters) had a correction greater than 1\%, and the largest correction was 35\% \footnote{Coincidence loss correction was not applied for the previous M31 UVIT catalog of \cite{2020ApJS..247...47L} so the brightest ($\lesssim$100) sources there require upward corrections $\sim$1\% to 35\%.}.
The conversion factors (UC) from counts rate to flux for each filter are given in \cite{2017Tandon2}, where updated values for UC can be calculated from the updated zero-point magnitudes given in Table 3 of \cite{2020AJ....159..158T}. 
We give these updated UC values in Table~\ref{table:confactors} here.
For convenience, Table~\ref{table:confactors} also gives the AB magnitude equivalent of a flux of $10^{-17}$erg cm$^{-2}$ s$^{-1}$ A$^{-1}$.
The uncertainty was calculated using the Poisson error in counts ($\sqrt{N_{src}+N_{bkg}}$, with $N_{src}$ the number of source counts and $N_{bkg}$ the number of background counts in the fitting box),  and given as a signal-to-noise ratio (S/N) for each source. 
Sources with S/N$\ge$3 were kept in a source catalog (with the remainder discarded as not sufficiently above noise).

\begin{figure}
\centering
\epsscale{1.0}
\plotone{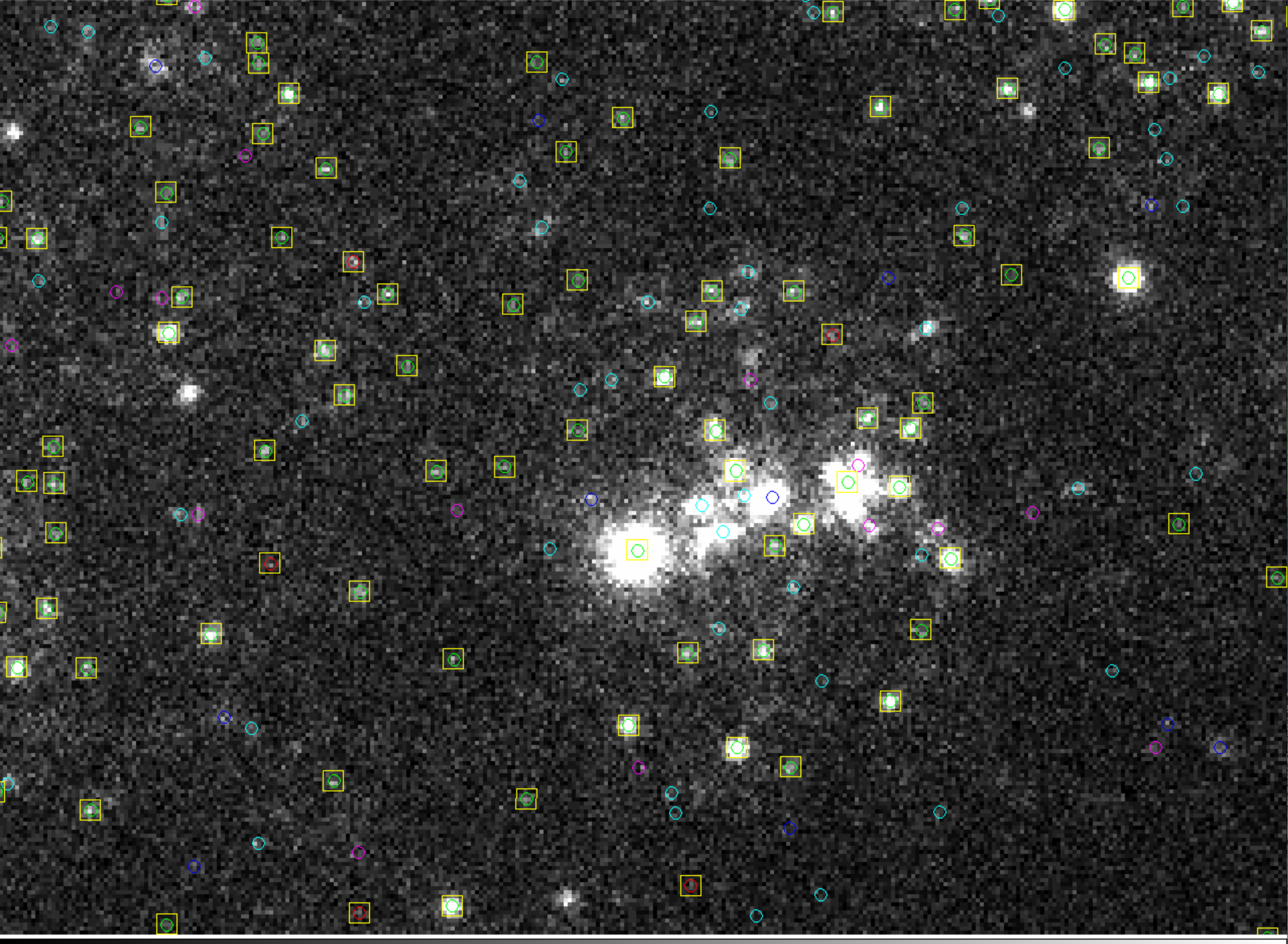}
\caption{A small region (2.2$^\prime$ wide by 1.5$^\prime$ high) in the field F1 showing the F148W image and the sources detected using CCDLAB.
Sources with eccentricity $e>0.8$ are circled with cyan; sources with $FWHM_Y>5$ circled with magenta: sources with $FWHM_X>5$ circled with blue; sources with $0.75<e<0.8$ are circled with red; the remainder circled with green.  The sources kept in the final selection have $e<0.8$, $FWHM_X<5$ and $FWHM_Y<5$ and are marked by yellow squares.
 \label{fig:check}
}
\end{figure}

The last step was visual examination of a large sample ($\sim1000$) of sources to determine that the Gaussian fits were giving reliable results. 
Field F1, which is the most crowded field and Field F7, which is a typical field in terms of source crowding were chosen for this, so both low crowding, intermediate crowding and high crowding regions were examined. 
We verified that the performance of the Gaussian fits did not depend on the waveband by comparing different wavebands then chose the F148W filter for more detailed examination. 
Many fits were of single compact sources, but others were found to be confused fits, usually caused by nearby ($<3^\prime$ separation) groups of two or more sources, where a single Gaussian was fit to more that group. 
The parameters of the confused fits were examined and found to fall into one of two categories: FWHM (full-width-at-half-maximum) in X (pixel RA coordinate) or Y (pixel Dec coordinate) $>5$ pixels ($2.1^{\prime \prime}$); or eccentricity $e>0.8$.
Thus the criterion used to accept a Gaussian fit was FWHM (full-width-at-half-maximum) in X (pixel RA coordinate) and in Y (pixel Dec coordinate) $<5$ pixels ($2.1^{\prime \prime}$); and eccentricity $e<0.8$.
Figure~\ref{fig:check} shows an example region (2.2$^\prime$  by 1.5$^\prime$) from the Field 1 F148W image with the source list before verification (marked by circles) and after verification (marked by yellow squares).
The image brightness scale was changed as needed to check whether the Gaussian fit to each source was acceptable. 
Overall the rejection rate was $\sim30$\% in uncrowded regions and  $\sim30$\% in crowded regions.

\section{Results}\label{sec:results}

A catalog was contructed for each of the six filter bands.
The numbers of sources for each filter band are listed in part (A) of Table~\ref{table:numbers}.  
The number of observed fields and the number of detected sources are given as two values: for sources with $>3\sigma$ flux measurements and for the subset with $>5\sigma$ flux measurements.

The F148W filter covers the majority of M31 (22 of 23 fields) and has the largest number of sources ($\simeq52,000$). 
The F172M filter (16 fields) covers $\sim$2/3 of M31, but has lower sensitivity compared to the other filters (see Table~\ref{table:confactors}) and has $\sim$13,000 sources.
The filters F169M (10 fields), N219M (10 fields) and N279N (9 fields) each cover $\sim$1/2 of M31
and have $\simeq22,000$, $\simeq11,000$  and $\simeq11,000$ sources, respectively. 
The larger number for the F169M filter is in part caused by longer exposure times for that filter and in part because the band is on the short wavelength side of the interstellar dust extinction peak at 220 nm. 
The F154W filter covers only 1 field and has $\simeq2500$ sources.

Because of the partial overlap of adjacent fields (see Figure~\ref{fig:fieldpositions}), some sources (of order 10\%) were detected twice (at different times), which we label 'duals'. 
Each overlap region between adjacent fields was examined to find these dual detections. 
For the duals, the separations between sources detected in the adjacent fields were found to be were within $1^{\prime \prime}$ of eachother, similar to the result shown in Fig. 6 of \citet{2020ApJS..247...47L}. 
Thus, duals were identified by matching sources from different fields in overlap regions with a maximum separation of $1^{\prime \prime}$. 
For the duals, the fluxes from the two different fields were compared. These show good agreement, as shown in Figure~\ref{fig:fluxcomp}, with expected deviations caused by the existence of variable sources.
The average flux and S/N were calculated for each dual using the total counts and exposure time from the two fields, and the duals were included in the final catalog presented here.
The numbers of duals is given in section (B) of Table~\ref{table:numbers}.

\begin{figure}
\centering
\epsscale{1.2}
\plotone{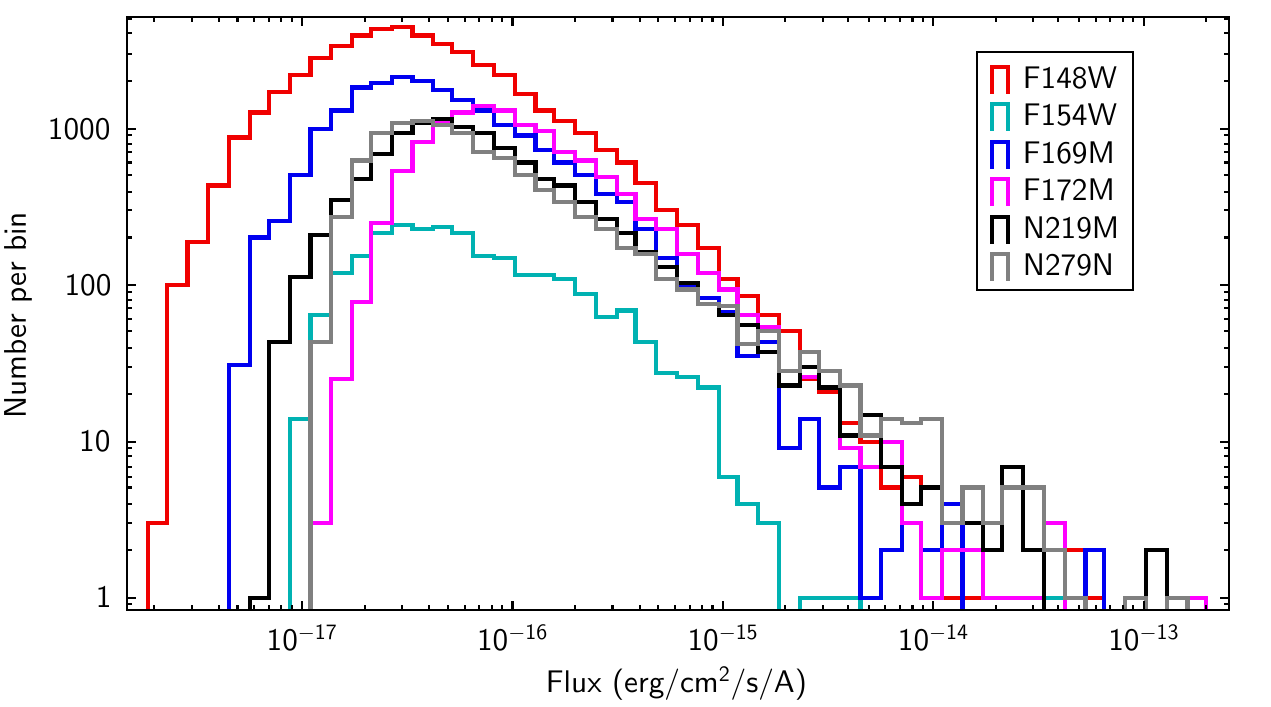}
\caption{The distribution of source fluxes for each of the filters: F148W, F154W,  F169M, F172M, N219M and N279N. 
 \label{MagDistn}
}
\end{figure}

The distributions of the fluxes of sources for each of the 6 UVIT filters are shown in Figure~\ref{MagDistn}. 
The measured numbers of sources increase to fainter fluxes  as expected because of increasing number of sources at fainter fluxes.
Then they decrease beyond a peak that because of the (variable) sensitivity limits of the observations.
The sensitivity limit of each image (field/filter combination) is different, in part because of
different exposure times, ranging from $\sim$3000 s to $\sim$30,000 s.
Source crowding, which varies greatly across M31, also reduces the detectability of faint sources,
thus, the flux detection limit varies across M31 as well as with filter band. 
We estimate the completeness limit of the survey by using the peak in the flux distributions for each filter. 
For F148W, the peak is at $\simeq 3.7\times10^{-17}$  erg  cm$^{-2}$ s$^{-1}$ A$^{-1}$, which translates to m$_{AB}\simeq$22.8. 
For F154W, F169M and F172M, the peaks are at $\simeq 4$, $\simeq 3$ and $\simeq 7.5\times10^{-17}$  erg  cm$^{-2}$ s$^{-1}$ A$^{-1}$, or m$_{AB}\simeq$22.6, 22.9 and 21.7, respectively.
For N219M and N279N, the peaks are at $\simeq 4.5$ and $\simeq 5 \times10^{-17}$  erg  cm$^{-2}$ s$^{-1}$ A$^{-1}$, or m$_{AB}\simeq$21.7 and $\simeq$21.4. 
The faintest detectable sources in each band are fainter than the incompleteness values given above, by an amount depending on the band.
 For F148W, the limit is at $\simeq 5\times10^{-18}$  erg  cm$^{-2}$ s$^{-1}$ A$^{-1}$ or m$_{AB}\simeq$25.0. 
For F154W, F169M and F172M, the limits are at $\simeq 1.2$, $\simeq 1.2$ and $\simeq 3\times10^{-17}$  erg  cm$^{-2}$ s$^{-1}$ A$^{-1}$ or m$_{AB}\simeq$24.0, 23.9 and 22.7, respectively.
For N219M and N279N, the limits are at $\simeq 2$ and $\simeq 3 \times10^{-17}$  erg  cm$^{-2}$ s$^{-1}$ A$^{-1}$ or m$_{AB}\simeq$22.6 and $\simeq$21.7. 

\begin{figure}
\centering
\epsscale{1.3}
\plottwo{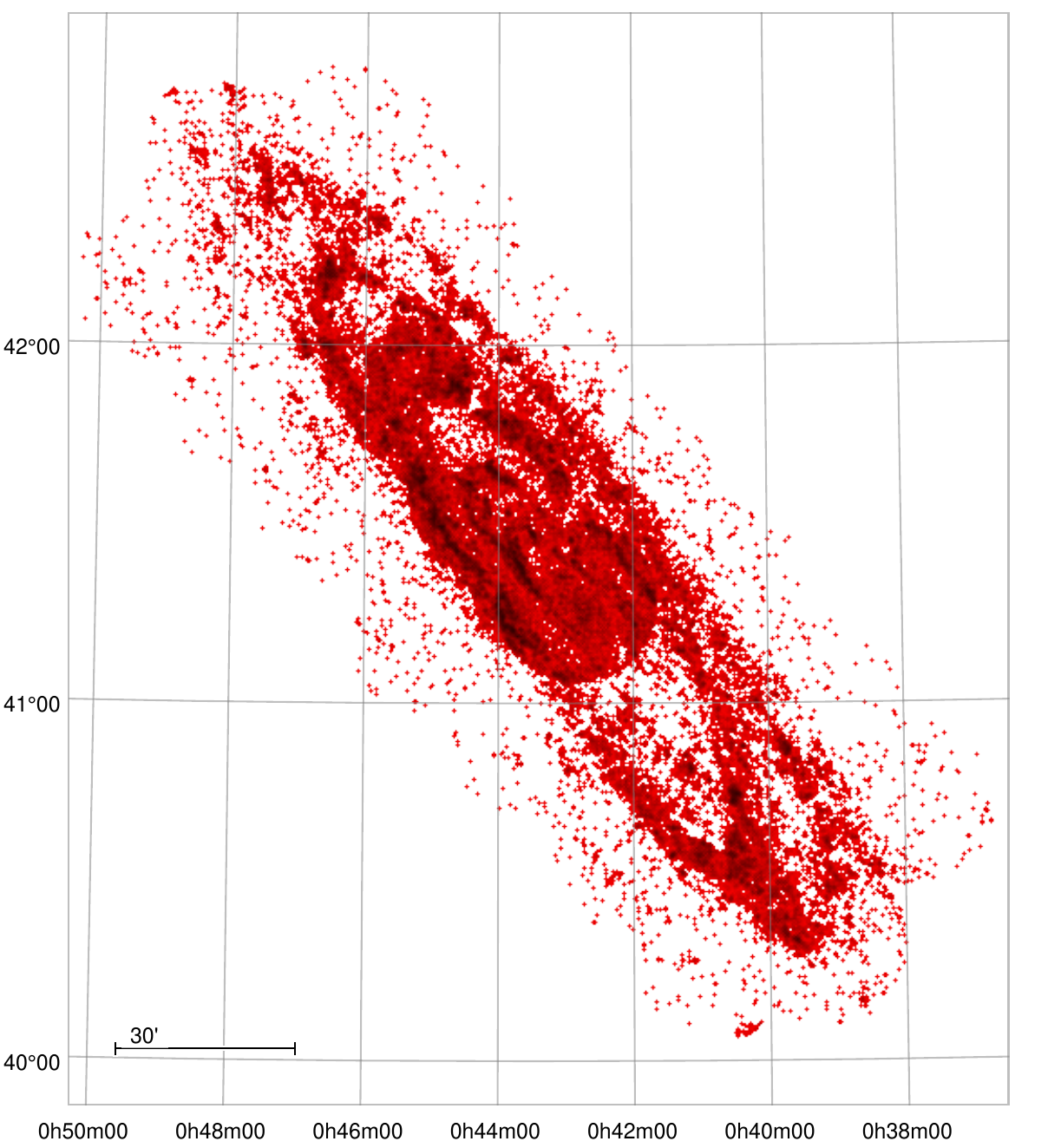}{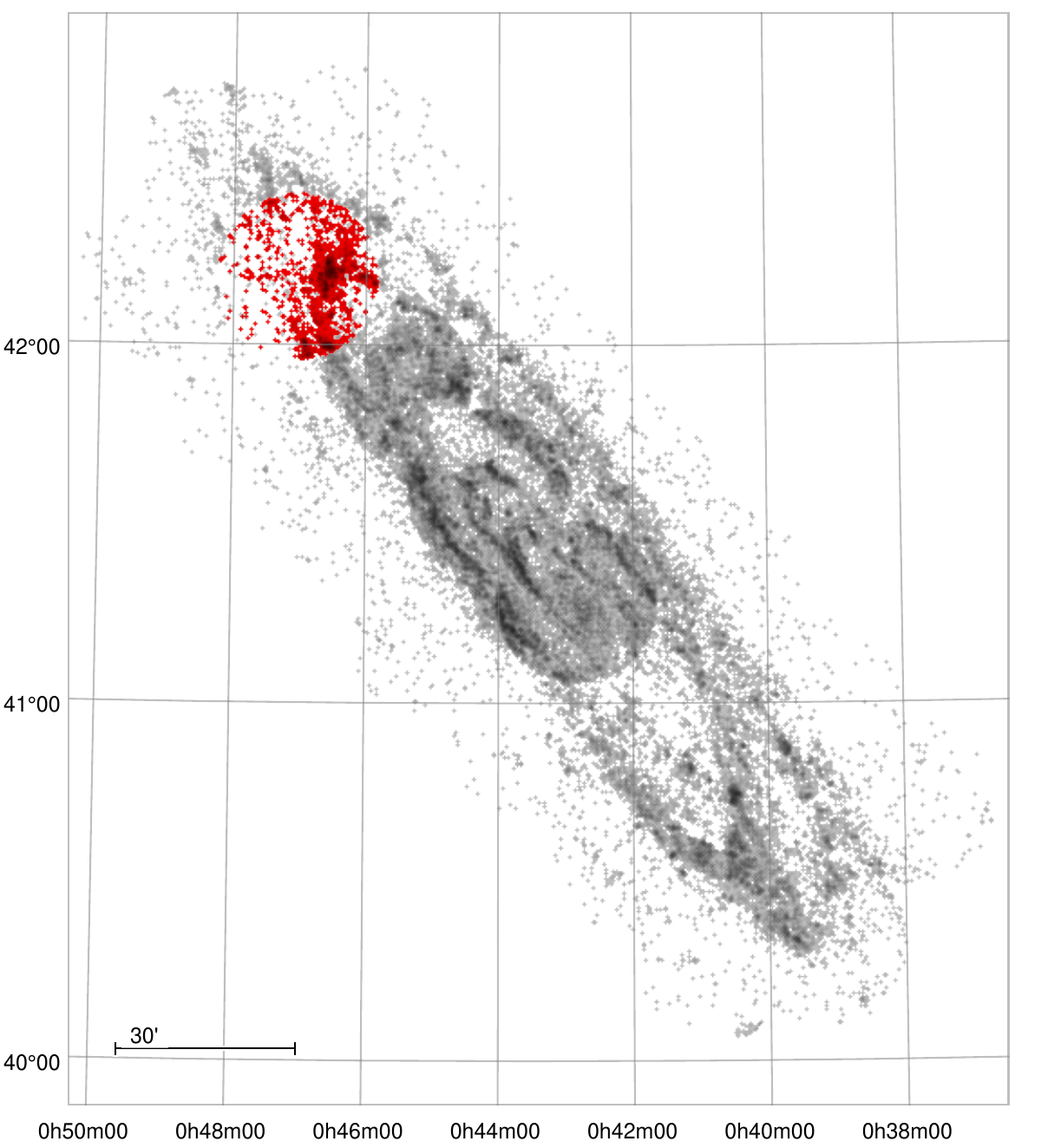}
\caption{Top panel: Map of UVIT F148W sources (red crosses) from the M31 survey.
Bottom panel: Map of UVIT F154W sources (red crosses) overlaid on the F148W sources (grey crosses).
 The sources shown have a minimum signal-to-noise (S/N) of 5.  \label{fig:F148F154map}
}
\end{figure}

\begin{figure}
\centering
\epsscale{1.3}
\plottwo{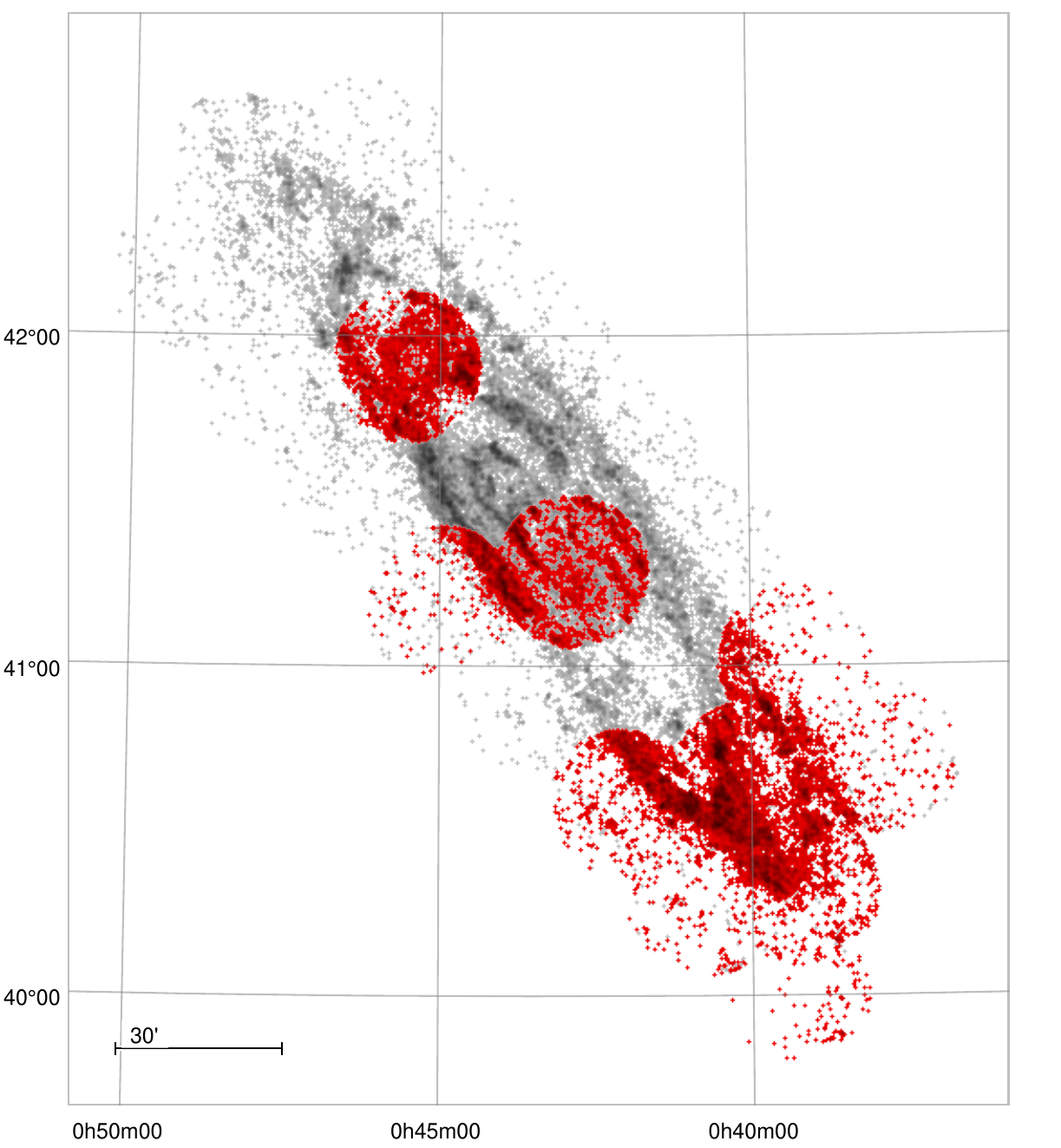}{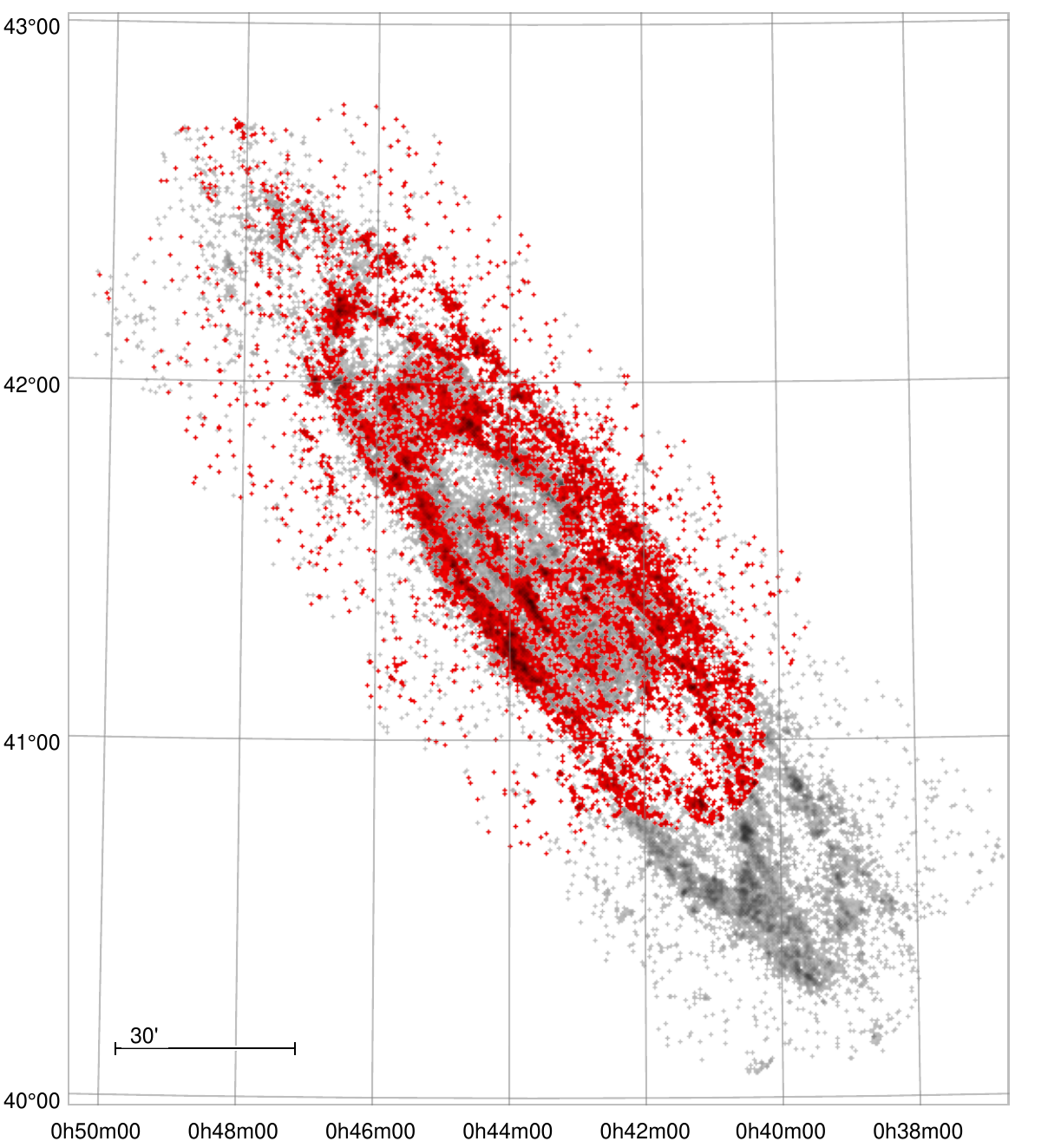}
\caption{Top panel: Map of UVIT F169M sources (red crosses) overlaid on the F148W sources (grey crosses).
Bottom panel: Map of UVIT  F172M sources (red crosses) overlaid on the F148W sources (grey crosses).
 The sources shown have a minimum S/N of 5.  \label{fig:F169F172map}
}
\end{figure}

\begin{figure}
\centering
\epsscale{1.3}
\plottwo{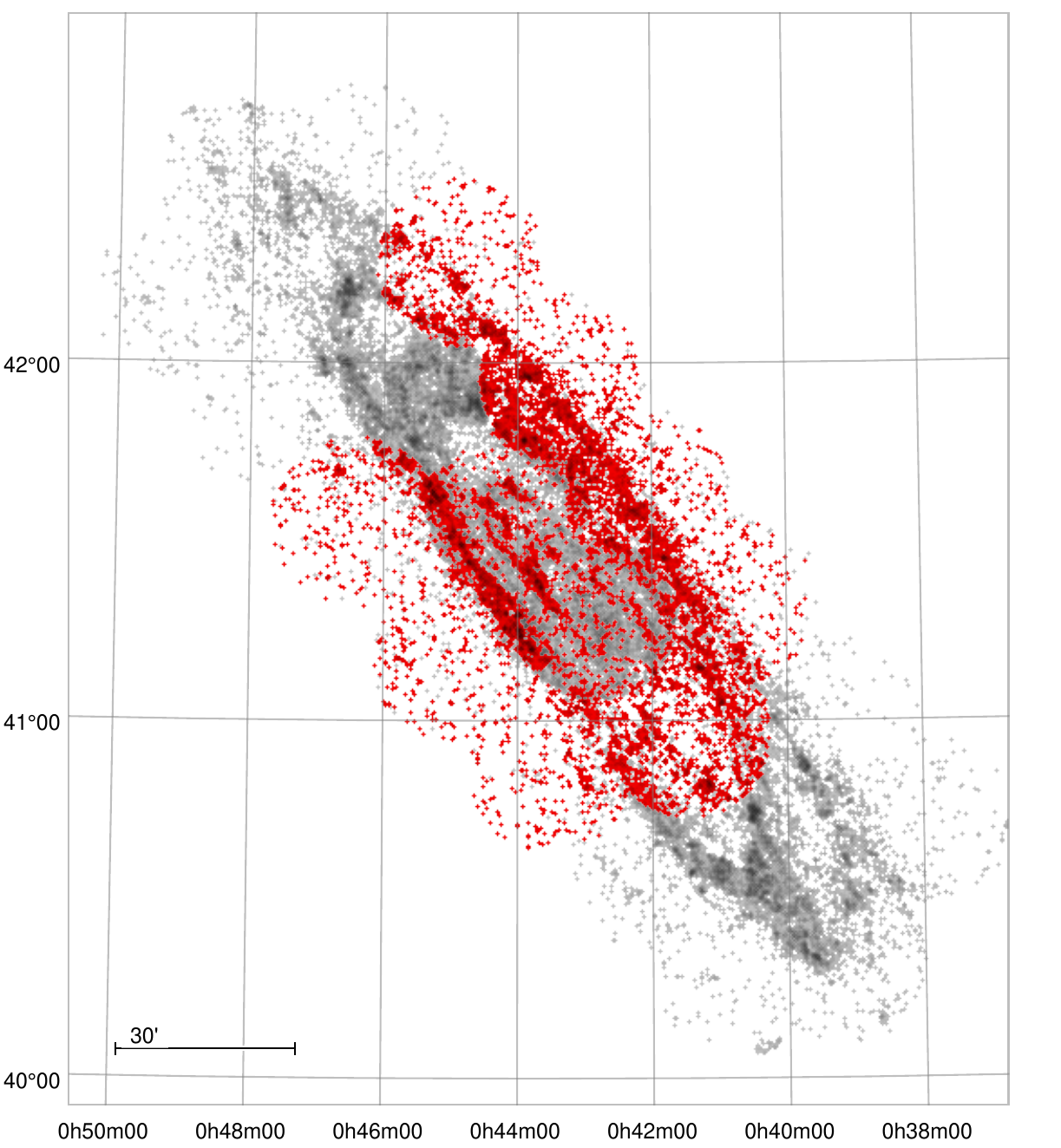}{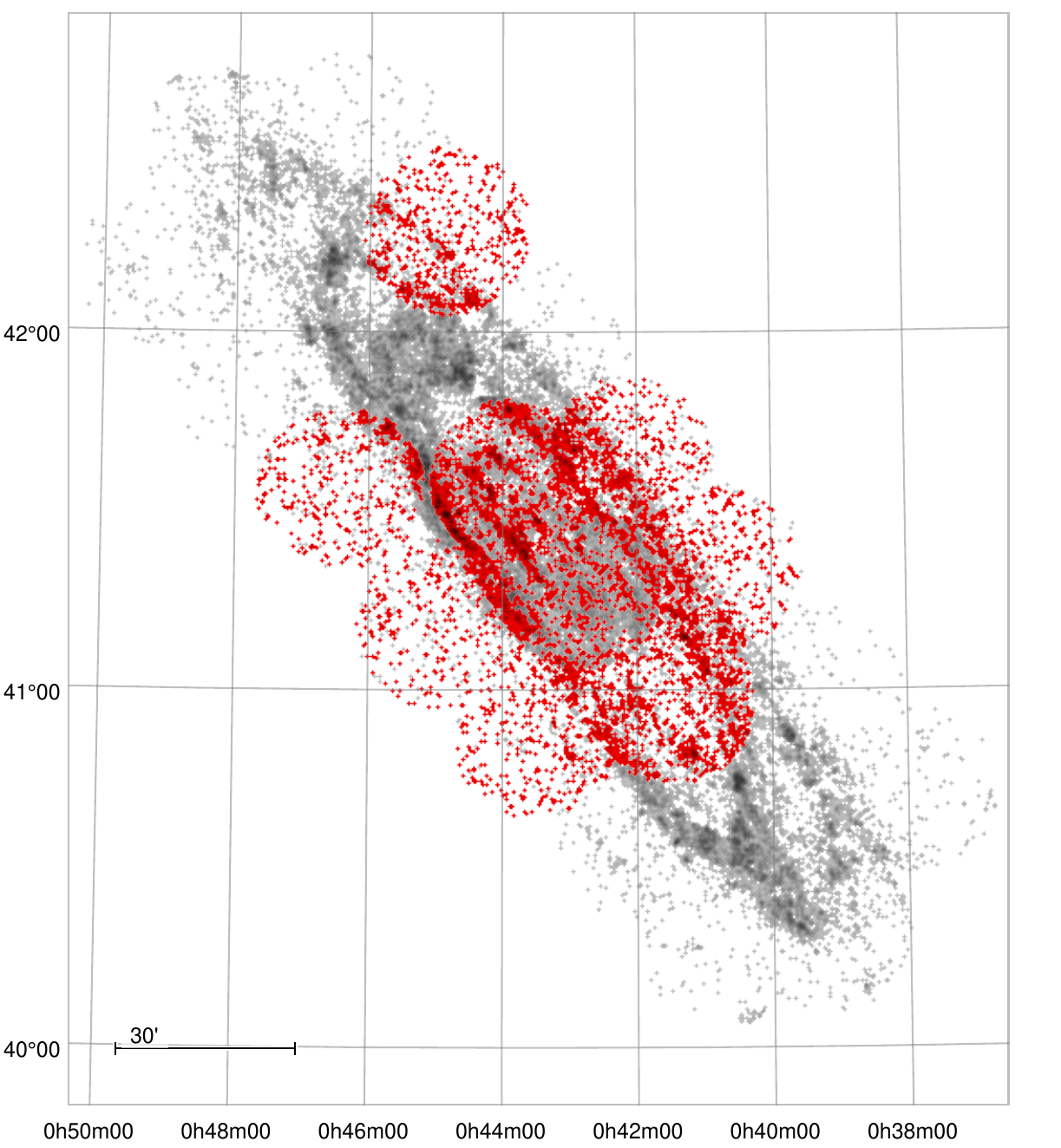}
\caption{Top panel: Map of UVIT  N219M sources (red crosses) overlaid on the F148W sources (grey crosses).
Bottom panel: Map of UVIT  N279N sources (red crosses) overlaid on the F148W sources (grey crosses).
 The sources shown have a minimum S/N of 5.  \label{fig:N219N279map}
}
\end{figure}

\begin{figure}
\centering
\epsscale{0.7}
\plotone{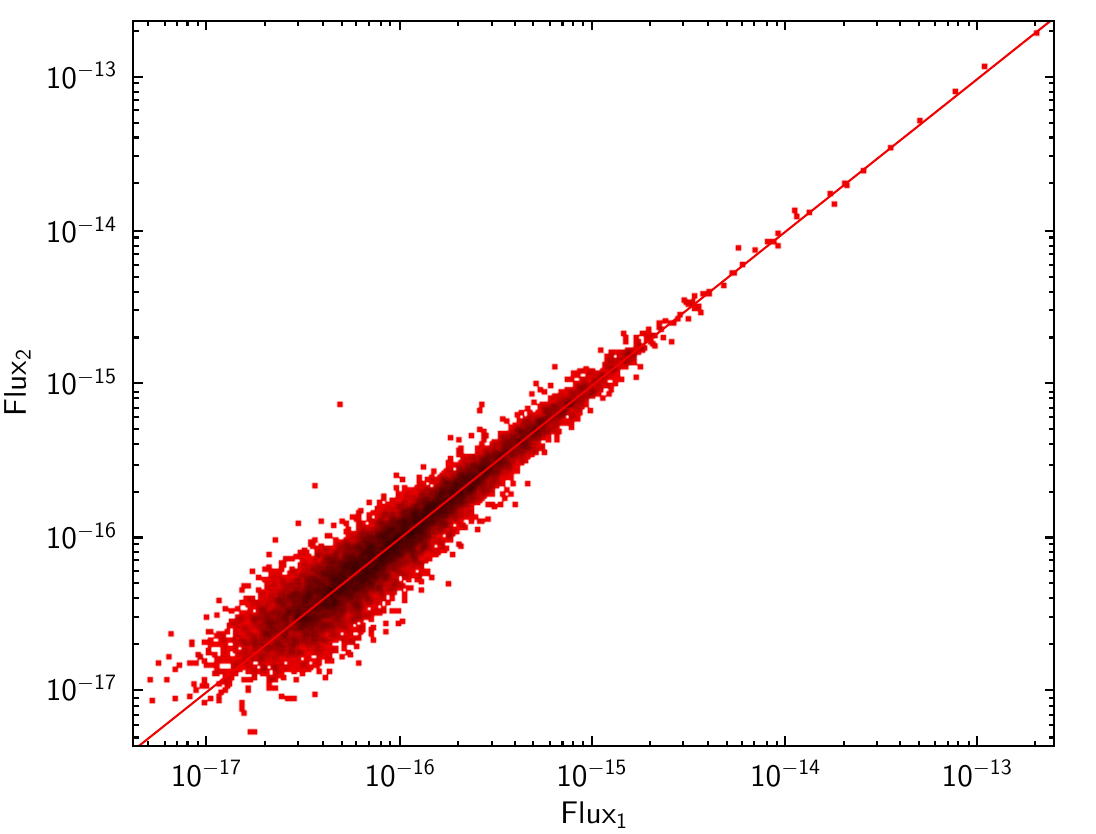}
\caption{
Flux comparison for dual sources: those detected twice in the same filter in the overlap region of different Fields (shown in Fig.~\ref{fig:fieldpositions}).
 \label{fig:fluxcomp}
}
\end{figure}

The positions of the F148W sources with SN$>$5 are shown in the top panel of Figure~\ref{fig:F148F154map}\footnote{ 
We chose to show the SN$>$5 sources because they are slightly less crowded than the SN$>$3 sources, although the appearance is much the same.}.
The spiral arms of M31 are seen as regions of high source density, with fields having higher exposure time (e.g. Fields 1, 2 and 8) showing more sources. 
The F154W sources (bottom panel, overlaid on the F148W sources in grey) generally coincide with the F148W sources in the single F154W observed Field 9. 
The F169M sources (top panel of Figure~\ref{fig:F169F172map}) are distributed similary to the F148W sources in the observed fields in southwest and center of M31, plus field 8.
The F172M sources  (bottom panel of Figure~\ref{fig:F169F172map}) have good coverage of M31 (all NE fields and central part of M31) and show prominently the spiral arms, but the bulge is less prominent than in the F148W and F169M filters.
The N219M and N279N sources  are shown in top and bottom panels of Figure~\ref{fig:N219N279map}, respectively, and cover the central regions of M31. The reason for this is that the NUV channel of UVIT failed before completion of the M31 UVIT survey. 
The spiral arms are prominent in both filters.
Foreground sources are a high fraction of the sources above and below the disk of M31, as shown by \cite{2020ApJS..247...47L}.

\subsection{Catalog Release}

The six catalogs, one for each filter (F148W, F154W, F169M,
F172M, N219M and N279N), were combined into a single combined M31 catalog, which is released with this paper. 
A sample of 10 entries from the combined M31 catalog is given in Table~\ref{table:uvittab}.
The columns are Filter, Field,  source position (J2000), flux, S/N, and AB magnitude. 

Here we analysed the data from the M31 UVIT survey to produce source catalogs
with $\simeq$115,000 sources detected in the UVIT FUV or NUV filters with S/N$\ge$3. 
The S/N includes statistical uncertainties, but the flux values also have systematic uncertainties.
The calibration uncertainties are $\simeq$1 to 2\% \citep{2020AJ....159..158T} for isolated source. 
There is a larger systematic uncertainty from source fitting in crowded regions which is $\sim$10\%.

This catalog will be useful for studies of stars and other sources in M31, especially those which have been measured in optical, infrared, radio and X-ray bands.
The combined table comprising the M31 source catalog is provided as an online attachment with this journal article.

\acknowledgments
This project is undertaken with the financial support of the Canadian Space Agency.
We thank the referee for recommendations which led to significant improvements.

\begin{deluxetable*}{llllll}
\tablecaption{M31 UVIT Observations\label{table:obs}  }
\tabletypesize{\scriptsize}
\tablehead{
\colhead{Field} &  \colhead{RA(deg)\tablenotemark{$a$}} & \colhead{Dec(deg)\tablenotemark{$a$}} & \colhead{Filter\tablenotemark{$b$}} & \colhead{Exposure Time} & \colhead{BJD0\tablenotemark{$c$}}         
}  
\startdata
1  & 10.71071 & 41.25023 & a$_1$,a$_2$,a$_3$\tablenotemark{$d$},c,  & 7736,17191,12632,10426         & 2457671(+0.856,+1134.256,+1512.147,+1133.850                        \\
 & & & d$_1$,d$_2$\tablenotemark{$d$},e,f & 3612,18056,7933,4347 & +1.194,+1135.223,+0.800,+1.140)   \\
2  & 11.03700 & 41.55735 & a$_1$,a$_2$\tablenotemark{$d$},d$_1$,d$_2$\tablenotemark{$d$},        & 7942,28395,7003,9353,                & 2457704(+0.135,+1463.979,+0.485,+1082.308,      \\
 & &  & e,f         & 7986,6120               & +0.135,+0.485)                    \\ 
3  & 10.34596 & 40.95242 & a,d,e,f         & 5006,10835,10900,5064               & 2458066(+0.447,+0.790,+0.790,+0.447)                                     \\
4  & 10.17567 & 41.32013 & a,d,e,f         & 4964,10277,10433,4989               & 2458067(+0.249,+0.462,+0.462,+0.249)                                     \\
5  & 10.49771 & 41.62544 & a,d,e,f         & 4981,10290,10522,5027               & 2458068(+0.600,+0.890,+0.890,+0.600)                                     \\
6  & 10.85308 & 40.85233 & a,d,e,f         & 5006,10272,10607,4773               & 2458070(+0.365,+0.789,)                                                     \\
7  & 11.22142 & 41.16111 & a$_1$,a$_2$\tablenotemark{$d$},c,d, & 4967,9528,9480,10692,   & 2458071(+0.187,+1815.788,+0.187,+0.390,             \\
 & & &  e, f  & 10793,3147 & +0.390,+0.187)   \\
8  & 11.36333 & 41.88303 & a,c,d            & 14925,11761,10491                    & 2458292(+0.309,+1222.048,+0.922)                                           \\
9  & 11.71750 & 42.20344 & a,b,d            & 7282,7582,8473                       & 2458356(+0.851,+1.131,+452.093)                     \\
10 & 10.83346 & 41.93324 & a,d,e            & 4981,10219,15340                     & 2458091(+0.149,+0.419,+0.149)                                          \\
11 & 11.18496 & 42.25849 & a,d,e,f         & 4972,10714,10806,5005               & 2458072(+0.060,+0.258,+0.258,+0.060)                                    \\
12 & 11.92225 & 41.85848 & a,d               & 4171,8869                             & 2458334(+0.252,+0.591)                                                     \\
13 & 11.59533 & 41.53595 & a$_1$,a$_2$\tablenotemark{$d$},d,e,    & 4974,6550,10706,10849,         & 2458092(+0.215,+1435.860,+0.433,+0.433,           \\
 & & & f  & 5009 & +0.215)   \\ 
14 & 9.993958 & 40.62872 & a,c               & 4728,10436                            & 2458426(+0.888,+1.042)                                                     \\
15 & 9.801542 & 40.28837 & a,c               & 7818,20650                            & 2458435(+0.076,+0.228)                                                     \\
16 & 10.51008 & 40.55778 & a,c               & 9655,20571                            & 2458428(+0.381,+0.786)                                                 \\
17 & 9.808417 & 40.9869  & a,c               & 4737,9997                             & 2458426(+0.009,+0.161)                                                    \\
18 & 9.476917 & 40.68349 & a,c               & 9490,18098                            & 2458434(+0.127,+0.280)                                                   \\
19 & 10.19363 & 40.25021 & a,c               & 4768,10429                            & 2458427(+0.583,+0.915)                                                \\
21 &  9.85417 & 39.98889 & c                   & 13664                                  & 2460287(+0.971)                                                              \\
23 & 12.09167 & 42.49444 & a,d               & 9080,8068                             & 2460314(+0.979,+1.865)                                          \\
24 & 12.33333 & 42.14722 & a,d               & 7748,9270                             & 2460316(+0.804,+1.690)                                          \\
25 & 11.54583 & 42.53889 & a,d               & 6056,6520                             & 2460318(+0.432,+1.719)                             \\
\enddata  
\footnotesize
\tablenotetext{a}{RA and Dec are the J2000 coordinates of the nominal pointing center of the observation.}
\tablenotetext{b}{Filter labels are a: F148W, b: F154W, c: F169M, d: F172M, e: N219M, f: N279N. If more than 1 observation then 1,2 etc. subscript is added.}
\tablenotetext{c}{BJD0 is the solar-system Barycentric Julian Date of the start of the observation. The common integer part for multiple observations is given as the first number.}
\tablenotetext{d}{Exposure times for the merged images are: F1 F148W image: 38299 s; F1 F172M image: 21789 s;  F2 F148W image: 36338 s; F2 F172M image: 16356 s; F7 F148W image: 14495 s; F13 F148W image: 11523 s.}
\end{deluxetable*}

\begin{deluxetable*}{ccccccc}
\tabletypesize{\scriptsize}
\caption{M31 Source Extraction Regions (Regions Of Interest, ROI) \label{table:roi} }
\startdata
& & & & & &  \\
Field &  Filter   & ROI x,y center & ROI radius &  Filter    & ROI x,y center & ROI radius \\ 
 &     & (pixels) & (pixels) &    &  (pixels) & (pixels) \\ 
\hline
1  & F148W     & 2435,2415& 1950   & F169M   &2408,2380 &	1992 \\
& F172M     &2415,2435&2040 & N219M   & 2261,2762 &2028 \\
   & N279N    & 2243,2752 &2025 \\
2  & F148W     & 2413,2458	&1955   & F172M   &2250,2496&1936\\
   & N219M   & 2267,2606&1988  & N279N    & 2857,2499	&1935 \\
3  & F148W     & 2342,2400&1953   & F172M   & 2377,2403&1942 \\
    & N219M   & 2310,2551&1979 & N279N    & 2287,2548&1950 \\
4  & F148W     & 2455,2400&1956   & F172M   &2424,2395&1948 \\
   & N219M   &2325,2689&1988  & N279N    & 2328,2531&1951 \\
5  & F148W     & 2493,2372&1942   & F172M   & 2545,2383&1947 \\
   & N219M   &2438,2505&1968 & N279N  &2391,2516&1962  \\
6  & F148W     & 2331,2386&1965  & F172M   &2357,2392&1951\\
   & N219M   &2270,2510&1997& N279N    & 2276	,2525&1968 \\
7  & F148W     & 2531,2533&2045   & F169M  & 2347,2413&2022 \\  
& F172M   &2221,2554&1948   & N219M   & 2189,2676&1974 \\
   & N279N    &2197,2652&1976 \\
8 & F148W     &2319,2012&1855  & F169M  & 2313,1954&1947 \\ 
& F172M     & 2368,2360&1962   & & \\
9  & F148W     & 2357,2363&1945   & F154W     &2363,2383&1945 \\
& F172M     & 2588,1972&1956   & & \\ 
10 & F148W     & 2400,2551&1942  & F172M   & 2409	,2548&1974 \\
   & N219M   &2281,2635&1962 \\
11 & F148W   &   2251,2509&2007   & F172M   & 2238,2519&2015\\
   & N219M   &2137,2666&2030 & N279N    & 2135,2641&2028 \\
12 & F148W     & 2416,2410&2015 & F172M   &2431,2426&2005 \\
13 & F148W     & 2546,2542&2005   & F172M   & 2406,2421&2000 \\
   & N219M   &2251,2517&2023   & N279N    & 2251,2477&2023 \\
14 & F148W     &2388,2393&2013   & F169M &2390,2395&2000 \\
15 & F148W     & 2441,2393&2025  & F169M & 2441,2367&2025 \\
16 & F148W     &2461,2363&1964  & F169M &2471,2436&1929 \\
17 & F148W     &2387,2406&1992  & F169M & 2370,2408&1995 \\
18 & F148W     & 2395,2383&1991   & F169M & 2428,2385&1982 \\
19 & F148W     & 2350,2499&1985 & F169M &2395,2494&1989 \\
21 & F169M & 2319,2377&2005  \\
23 & F148W     & 2345,2436&2000   & F172M & 2365,2436&1995 \\
24 & F148W     & 2411,2406&1995   & F172M & 2395,2507&2010 \\
25 & F148W     & 2309,2456&1995   & F172M & 2301,2403&2007 \\
\hline
\enddata  
\footnotesize
\end{deluxetable*}

 \begin{table*}
\caption{UVIT Filters, UC$^{a}$ (erg  cm$^{-2}$ s$^{-1}$ A$^{-1}$/(count s$^{-1}$)) and Flux to AB Magnitude Conversions \label{table:confactors}}
\centering
\begin{tabular}{ccccccc}
\toprule
Filter & $ \lambda_{\textrm{mean}}$ (\r{A})$^{b}$  & $ \Delta \lambda$ (\r{A})$^{b}$  & UC & UC error &  mAB$^{c}$   \\
\hline
F148W & 1481 & 500 & 2.864$\times10^{-15}$ &2.7$\times10^{-17}$ & 24.24\\
F154W & 1541 & 380 & 3.571$\times10^{-15}$ &4.0$\times10^{-17}$& 24.15\\
F169M & 1608 & 290 & 4.574$\times10^{-15}$ & 3.9$\times10^{-17}$& 24.06\\
F172M & 1717 & 125 & 1.142$\times10^{-14}$ & 1.7$\times10^{-16}$&23.92\\
N219M & 2196 & 270 & 4.920$\times10^{-15}$& 7.7$\times10^{-17}$& 23.38\\
N279N & 2792 & 90  & 3.790$\times10^{-15}$ &3.8$\times10^{-17}$&22.86\\
 \hline
\end{tabular}
\footnotesize
\tablenotetext{a}{UC (CPS to Flux Conversion Factors) are based on the updated zero-point magnitudes given in \cite{2020AJ....159..158T}.} 
\tablenotetext{b}{The mean wavelength,  $\lambda_{\textrm{mean}}$, and the bandwidth,  $\Delta \lambda$,  are from \cite{2017Tandon2} Table 3.}
\tablenotetext{c}{AB magnitude equivalent of a flux of $10^{-17}$erg cm$^{-2}$ s$^{-1}$ A$^{-1}$.} 
\end{table*}

 \begin{table*}
\caption{Summary of M31 UVIT Source Numbers\label{table:numbers}}
\centering
\begin{tabular}{crlccrlcc}
\toprule
A. Filter & \# of Fields & \# of Sources$^{a}$ &Sky Area &  Filter & \# of Fields & \# of Sources$^{a}$ & Sky Area \\
          &              &                     & (deg$^2$) &          &               &             & (deg$^2$)\\          
\hline
F148W & 22 &  53815(44995) & 3.52 & F154W & 1 &2486(1988)  & 0.18 \\  
F169M & 10 &  22850(18368) & 1.61 & F172M & 16 & 13547(11289) & 2.46  \\  
 N219M & 10 & 11493(10224) & 1.61 & N279N &  9& 11094(7725)  &  1.46 \\ 
\hline
B$^{b}$. Filter & & \# of Sources& &  Filter  &  &  \# of Sources \\
\hline
F148W  & &4968  & &  F154W& &  0 &  \\  
F169M & &1842 & &  F172M && 1530 &   \\  
N219M & &696 &   & N279N& &984 &  \\ 
 \hline
\hline
\end{tabular}
\footnotesize
\tablenotetext{a}{Numbers of sources detected at signal-to-noise S/N$\ge3$. Numbers of sources detected at S/N$\ge5$ are in brackets. 'Duals' are  included in these numbers. 
} 
\tablenotetext{b}{'Duals' are the same source observed in two adjacent fields in their overlap areas.}    
\end{table*}

 \begin{table*}
\caption{10 Entries  from the Combined Catalog\label{table:uvittab}}
\centering
\begin{tabular}{lccccccccc}
\toprule
Filter &Field&RA &Dec& F$_{\lambda}$&S/N&$m_{AB}$ \\
&  &($^\circ$ J2000)   &   ($^\circ$ J2000)  &(erg/cm$^{2}$/s/A) &&    \\
\hline
F148W&	F1&	10.410703&	41.289005&	1.55E-17&	9.9&	23.76\\
F148W&	F1&	10.410862&	41.285318&	7.71E-18&	6.5&	24.52\\
F148W&	F1&	10.411274&	41.287788&	2.24E-17&	12.1&	23.36\\
F148W&	F1&	10.411623&	41.283694&	3.96E-17&	15.9&	22.75\\
F148W&	F1&	10.411755&	41.291828&	7.54E-18&	6.3&	24.55\\
F148W&	F1&	10.412639&	41.273101&	3.21E-18&	3.9&	25.47\\
F148W&	F1&	10.413233&	41.315996&	1.73E-17&	10.2&	23.64\\
F148W&	F1&	10.413372&	41.297675&	4.24E-18&	4.6&	25.17\\
F148W&	F1&	10.413438&	41.304621&	9.83E-18&	7.8&	24.26\\
F148W&	F1&	10.413477&	41.254122&	8.74E-18&	7.4&	24.39\\
\end{tabular}
\footnotesize
$\quad$\\
\end{table*}

\end{document}